\documentclass[aps,prb,reprint,longbibliography]{revtex4-1}

\usepackage{graphicx}
\usepackage{dcolumn}
\usepackage{bm}
\usepackage{color, soul}
\usepackage{xcolor}
\usepackage{amsmath, amssymb}

\begin{document}

\title{Tailoring spin wave channels in a reconfigurable artificial spin ice}

\author{Ezio Iacocca}
\affiliation{Department of Mathematics, Physics, and Electrical Engineering, Northumbria University, Newcastle upon Tyne, NE1 8ST, UK}
\affiliation{Department of Applied Mathematics, University of Colorado, Boulder, Colorado 80309-0526, USA}

\author{Sebastian Gliga}
\affiliation{Swiss Light Source, Paul Scherrer Institute, 5232 Villigen, Switzerland}
\affiliation{School of Physics and Astronomy, University of Glasgow, Glasgow G12 8QQ, United Kingdom} 

\author{Olle G. Heinonen}
\affiliation{Materials Science Division, Argonne National Laboratory, Lemont, Illinois 60439, USA}
\affiliation{Center for Hierarchical Materials Design, Northwestern-Argonne Institute for {Science} and Engineering, Evanston, Illinois 60208, USA}

\begin{abstract}
Artificial spin ices are ensembles of geometrically-arranged, interacting nanomagnets which have shown promising potential for the realization of reconfigurable magnonic crystals. Such systems allow for the manipulation of spin waves on the nanoscale and their potential use as information carriers. However, there are presently two general obstacles to the realization of artificial spin ice-based magnonic crystals: the magnetic state of artificial spin ices is difficult to reconfigure and the magnetostatic interactions between the nanoislands are often weak, preventing mode coupling. We demonstrate, using micromagnetic modeling, that coupling a reconfigurable artificial spin ice geometry made of weakly interacting nanomagnets to a soft magnetic underlayer creates a complex system exhibiting dynamically coupled modes. These give rise to spin wave channels in the underlayer at well-defined frequencies, based on the artificial spin ice magnetic state, which can be reconfigured. These findings open the door to the realization of reconfigurable magnonic crystals with potential applications for data transport and processing in  magnonic-based logic architectures. 
\end{abstract}

\maketitle

\section{\label{sec:Intro}Introduction}

{In magnonics~\cite{Kruglyak2010,Lenk2011,Demokritov2013,Chumak2015}, spin waves are utilized as carriers of information for a wealth of technologically relevant functions. The most straight-forward approach consists in transporting microwaves via low-loss insulating magnetic materials such as YIG~\cite{Serga2010}. From this basic idea, the field has expanded in {multiple} technologically relevant directions. One avenue of research has explored the design of logic gates based on spin-wave interferometry~\cite{Hertel2004,Chumak2015} and the directional propagation of spin waves~\cite{Vogt2012}. Another consists in patterning the magnetic material as a superlattice to support spin-wave band structure, which exhibits band gaps of prohibited propagating energy bands~\cite{Nikitov2001}. Such patterned magnetic materials, called ``magnonic crystals'' in analogy to photonic crystals, were for example used to realize a magnonic transistor~\cite{Chumak2013}. Furthermore, intense research has focused on reconfigurable~\cite{Grundler2015} band structures based on approaches that manipulate the energy landscape~\cite{Karenowska2012,Obry2013,Wang2017}, the magnetic ground state~\cite{Tacchi2011,Gubbiotti2018} as well as current-induced chiral magnetization states~\cite{Sprenger2019}. The advantage of a reconfigurable lattice is the active modification of its band structure, e.g. by toggling or manipulating the existence of a magnonic band gap. The possibility of defining myriad functions using magnonic crystals makes them attractive for all-magnetic logic and computing applications that can miniaturize microwave circuitry~\cite{Grundler2015}.}

\begin{figure}[b]
\centering \includegraphics[width=3.3in]{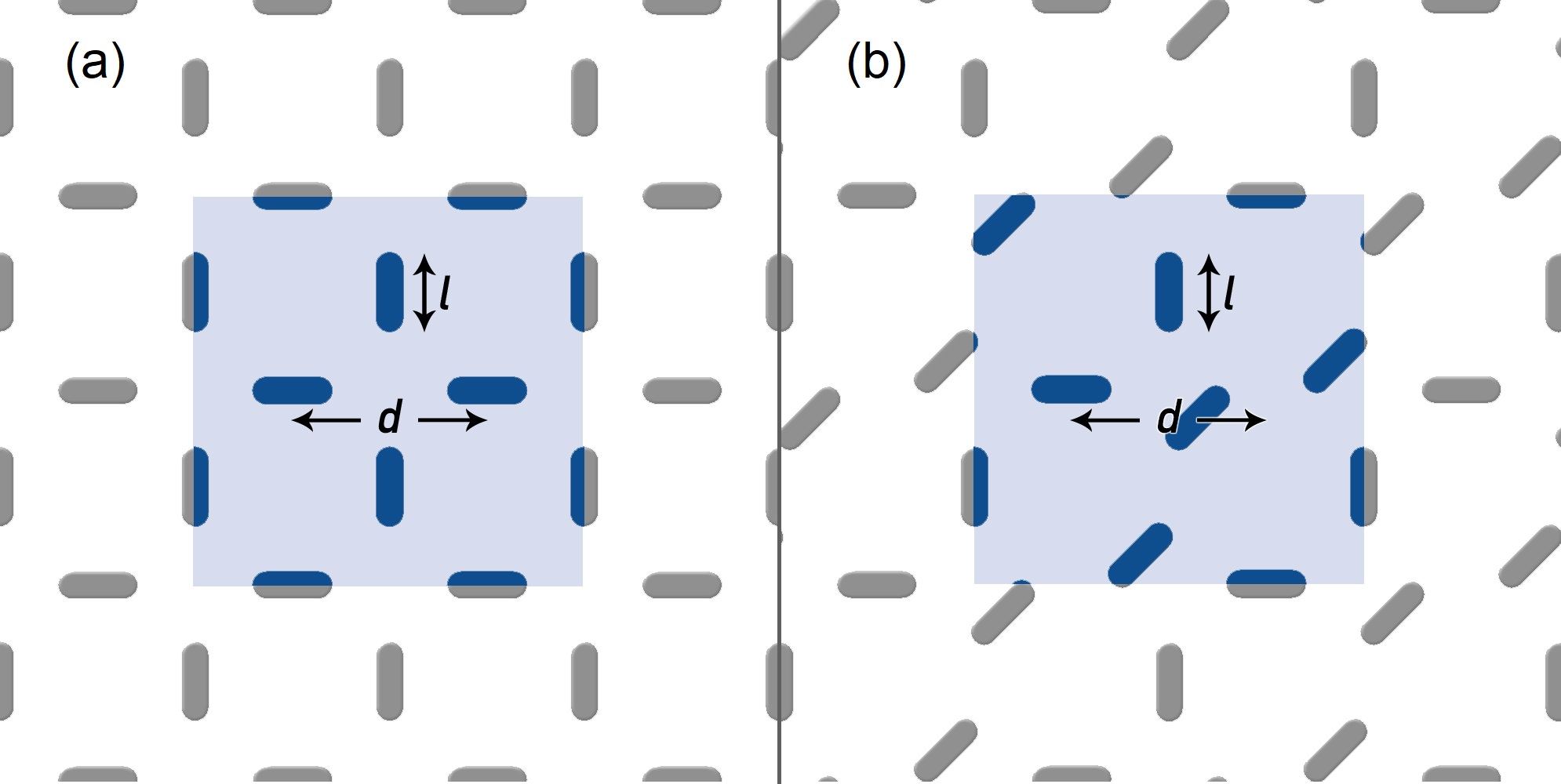}
\caption{ \label{fig1} Geometries of (a) square ice and (b) charge ice. The respective simulated {supercells} are highlighted in blue. }
\end{figure}
In this context, artificial spin ices have shown potential as reconfigurable superlattices~\cite{Heyderman2013}. Artificial spin ices were originally designed to mimic the frustration found in rare earth Pyrochlore compounds, thereby creating geometric frustration by design~\cite{Wang2006}. Hence, artificial spin ices have been typically studied in the context of massively degenerate ground states, geometric frustration, and constrained disorder~\cite{Nisoli2017}. Recently, it has become clear that artificial spin ices show promise for creating functional materials, such as ratchets~\cite{Gliga2017}, reprogrammable superconducting flux diodes~\cite{Wang2018} or logical gates~\cite{Arava2018,Arava2019}. In particular, {the resonant dynamics can be tailored based on the magnetic structure in these systems~\cite{Tacchi2011,Gliga2013,Gliga2015,Iacocca2016,Jungfleisch2016,Bhat2016,Zhou2016,Iacocca2017c,Bang2019,Dion2019,Lendinez2019,Ghosh2019,Arroo2019}.} Recently, nanomagnets subject to interfacial Dzyaloshinskii-Moriya interaction have been shown to give rise to topologically nontrivial magnon bands~\cite{Iacocca2017c} as well as to comprise building blocks for complex synthetic magnetic states and structures, including an artificial kagome lattice~\cite{Luo2019}. {The realization of magnonic crystals that exhibit topological edge modes based on these chiral nanomagnets would represent a promising research direction in this field.}

However, a shortcoming of artificial spin ices so far is that desired global magnetic states can {be only} achieved through careful thermal annealing or demagnetizing field protocols~\cite{morgan2011,Farhan2013,Sklenar2019}, while local magnetic states can be achieved by manipulating the state of individual nanomagnets using a magnetized tip~\cite{Gartside2018,Lehmann2019}. In order to achieve practical reconfigurable systems, global as well as local magnetic states should be accessible through simple and well-defined protocols. 

Recent efforts in this direction have led to the creation {of} a fully reconfigurable artificial spin ice~\cite{Wang2016,Wang2018}. This system, called magnetic charge spin ice~\cite{Wang2016} (charge ice), reproduces the magnetic volume charge distribution {of} the artificial square ice (square ice), while introducing diagonal elements that enable straight-forward field control of the magnetic state. {In particular, Wang et al. reported a 99\% yield in achieving single-type long-range order in the charge ice {in Ref.~\onlinecite{Wang2016}}. Such a reproducible high yield is essential to predictably control the magnon modes in a practical magnonic crystal. Deviations from such modes, e.g., caused by defects or disorder, may lead to the quenching or enhancement of spectral features~\cite{Gliga2013} as well as the appearance of localized modes.}

Schematics of the square and charge ices are shown in Fig.~\ref{fig1}a and b, respectively. In order to replicate the magnetic volume charge distribution in square ice, the  charge ice is subject to a {geometric} constraint between the length, $l$, of the nanoislands and the {center-to-center inter-nanoisland distance of a square ice,} $d$:
\begin{equation}
  d = l\left(1+\sqrt{2}\right).
  \label{eq:geometry}
\end{equation}
This constraint results in a relatively large distance $d$, leading to a weak dipolar coupling between the nanoislands. In the context of magnonics, such a weak coupling limits the effect of the lattice periodicity on the resulting spin wave band structure. Essentially, the spin wave band structure is that of the bulk and edge modes of a single, isolated nanoisland~\cite{Carlotti2014} and the advantage of using an artificial spin ice as a reconfigurable magnonic crystal is lost.

In order to enhance the coupling between the nanomagnets while respecting the geometric constraint of charge ice, we consider an artificial spin ice on top of a soft magnetic thin film. This {thin film} underlayer  couples to the individual nanoislands, mediating magnetic interactions between different nanoislands while allowing the magnetization of each nanoisland to be reconfigurable. To achieve such a balance, we consider a soft magnetic underlayer separated from the nanoislands by a thin non-magnetic layer, such that the interlayer exchange interaction between the magnetization in the underlayer and the nanoislands is weak. In this way, the interactions between the spin ice and the underlayer give rise to spin wave modes supported by the underlayer, providing an additional dynamic coupling between the nanoislands. {This is similar to the spin wave hybridization recently observed experimentally in magnetic ellipses patterned on top of a Pt-buffered Permalloy underlayer~\cite{Graczyk2018}.}
\begin{figure}[t]
\centering \includegraphics[width=2.5in]{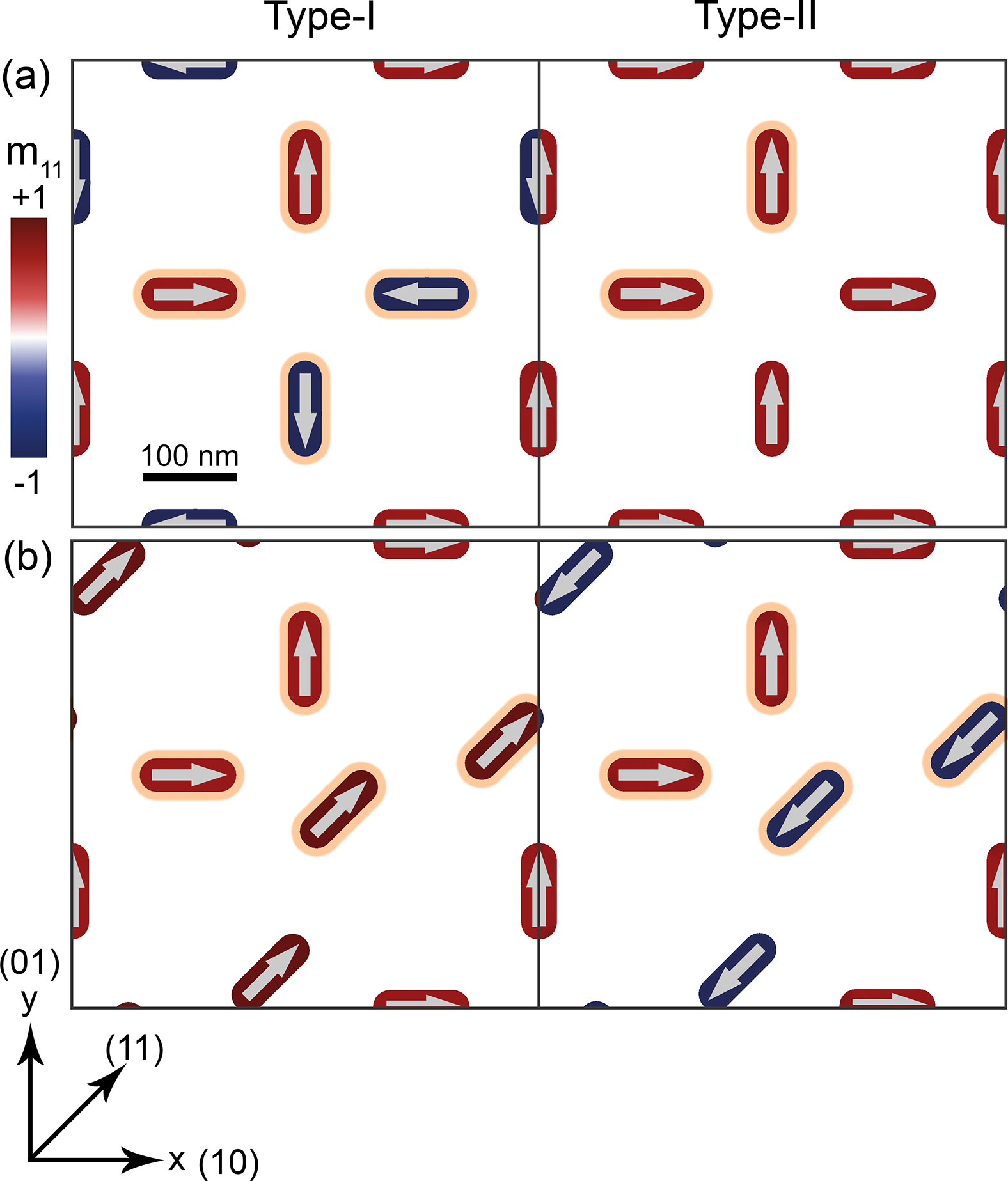}
\caption{ \label{fig:ice_types} Magnetic states in the simulated {supercell} of (a) square ice and (b) charge ice. Both Type-I and Type-II states are depicted. The arrows schematically indicate the direction of the magnetization in each nanoisland while the color map codes the magnitude of the projection of the magnetization along the (11) diagonal. {The nanoislands highlighted in {orange} shade compose the unit-cell {in} each case.} }
\end{figure}

{In addition, it is essential to keep well-defined magnetic states accessible, e.g. through magnetic field protocols. While one advantage of artificial spin ices is the possibility of accessing {a multitude of} different states, the stable state of a soft magnetic underlayer may depend on the reconfiguration history. To mitigate this issue, }we impose a preferential orientation of the magnetization in the underlayer by applying an exchange bias~\cite{Nogues1999}. The exchange bias is strong enough so that the underlayer equilibrium magnetization is not significantly disturbed when the magnetization in the nanoislands is reconfigured -- and at the same time sufficiently weak that it does not pin the magnetization in the {underlayer}. 

Based on micromagnetic simulations, we investigate the eigenmodes of square and charge ices coupled through a soft magnetic underlayer.  We find that the imposition of coupling a uniaxially anisotropic magnetic thin film to the spin ice lattice leads to interesting and complex modes. Indeed, the eigenmode spectrum of the coupled system is a combination of dipolar-dominated modes favored in the artificial spin ice and exchange-dominated dynamics favored in the underlayer. {The artificial spin ice lattice establishes standing spin-wave modes in the uniaxially anisotropic magnetic thin film. Remarkably, some of those modes are reminiscent of extended spin wave channels present in antidot magnonic crystals~\cite{Neusser2011,Sklenar2013} while others are spatially localized. The system thus acts as a magnonic waveguide (the underlayer), which can be toggled to support or inhibit energy propagation by an array of nanoantennas (the artificial spin ice). This feature may have potential applications for data propagation and manipulation~\cite{Chumak2017} as they are highly anisotropic and depend on the configuration of the artificial spin ice.}

\section{Results and discussion\label{sec:results}}
\subsection{Square ice and charge ice {dynamics}\label{subsec:square_and_charge}}

As described in the Introduction, the constraint Eq.~(\ref{eq:geometry}) for building a charge ice leads to a large inter-nanoisland separation, such that the nanoislands are weakly coupled. We consider throughout this study nanoislands made of Permalloy (Py) with saturation magnetization $M_{\rm S}=790$~kA/m, exchange constant $A=13$~pJ/m and negligible magnetocrystalline anisotropy.  The nanoislands have width $w=35$~nm, thickness $t=15$~nm, and length $l=100$~nm. The nanoisland spacing is then of $d\approx240$~nm following Eq.~\eqref{eq:geometry}.

We consider two different magnetization states. In square ices, shown in Fig.~\ref{fig:ice_types}(a), the ground state is labeled Type-I. 
This configuration has been {experimentally} shown to {be only} reachable through careful thermal annealing~\cite{morgan2011,Farhan2013}. The Type-II is a remanent state that can be {experimentally} obtained by applying a {saturating} external field {e.g.} along the (11) diagonal direction, and then letting the magnetization relax {as the field is removed}. The corresponding charge ice Type-I and Type-II configurations are shown in Fig.~\ref{fig:ice_types} (b). In contrast to the square ice, the charge ice can {be experimentally reconfigured in a relatively easy manner} between Type-I and Type-II by applying {an external field on the order of $0.1$~T} along the {(11) and ($\bar{1}\bar{1}$)} directions, respectively{, as demonstrated in Refs.~\onlinecite{Wang2016,Wang2018}. Here, we simulate each configuration by setting an initial magnetization state according to Fig.~\ref{fig:ice_types} and then finding the energy minimum with the protocol described in Appendix B.}
\begin{figure}[t]
\centering \includegraphics[width=3.2in]{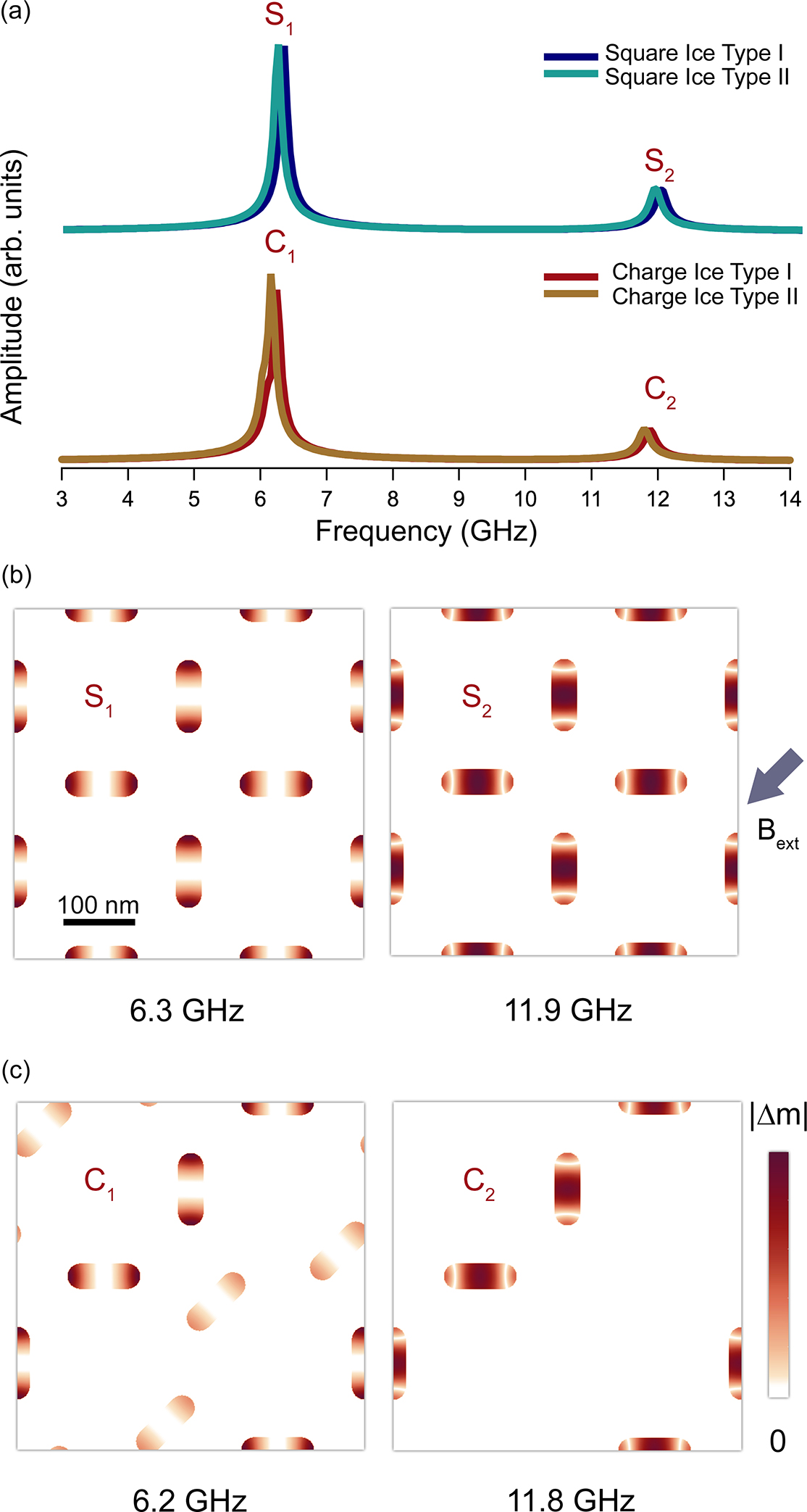}
\caption{ \label{fig:square_ice_I} (a) Spectra of square and charge {ices} in the Type-I and Type-II configurations{, following perturbation by a 50 ps field pulse, $B_\mathrm{ext}$, along the ($\bar{1}\bar{1}$) direction indicated by the large arrow}. The spatially-resolved modes {obtained for Type-I artificial spin ices} are shown in (b) for square ice and {in} (c) for charge ice. In both cases, the lowest-frequency mode corresponds to an edge mode and the highest-frequency mode corresponds to a bulk mode. {The diagonal elements in the charge ice are unperturbed owing to the direction of the applied field pulse. The colorscale indicates the mode amplitude.} }
\end{figure}

{Even though the charge ice is inspired by the charge configuration of a square ice, the primitive cells (or unit cells) that span each lattice for a given long-range magnetic order are different. The nanoislands forming the unit cell for both artificial spin ices in Type-I and Type-II configurations are indicated by an {orange} shade in Fig.~\ref{fig:ice_types}. In a square ice, four nanoislands form the unit cell in the Type-I state, while only two nanoislands are required in the Type-II state. In the charge ice, both Type-I and Type-II states have unit cells composed of four nanoislands. This difference is due to the symmetry of the square ice, in which the smallest unit cell is that of the Type-II configuration.}

The spectra for Type-I and Type-II spin ices are shown in Fig.~\ref{fig:square_ice_I}(a). {Here, we simulated the micromagnetic supercell imposing periodic boundary conditions in the (x,y) plane. First, the magnetic energy for each magnetization state without external field was minimized using the protocol described in Appendix~\ref{app:ground}. Then, dynamics was excited with a short (50 ps) magnetic field pulse along the ($\bar{1}\bar{1}$) direction, as described in Appendix~\ref{app:spinwave_ASI}}. Both the square and charge ices exhibit essentially the same response, independently of their  magnetization state, as a result of the weak coupling between nanoislands.

In Fig.~\ref{fig:square_ice_I}(b), the spatially-resolved square ice modes S$_1$ ($6.3$~GHz) and S$_2$ ($11.9$~GHz) are shown. The spatially-resolved modes in the charge ice, C$_1$ ($6.2$~GHz) and C$_2$ ($11.8$~GHz), are shown in Fig.~\ref{fig:square_ice_I}(c). In both cases, the lowest-frequency mode corresponds to an edge mode {(with the largest amplitude)} and the highest-frequency mode corresponds to a bulk mode, in agreement with earlier studies~\cite{Gliga2013,Jungfleisch2016,Iacocca2016}. We note that the diagonal elements in the charge ice are essentially unperturbed because the perturbation field was applied along the ($\bar{1}\bar{1}$) direction.

\subsection{\label{subsec:single_island}{Dynamics of a} single nanoisland on underlayer}

\begin{figure}[t]
\centering \includegraphics[width=3.2in]{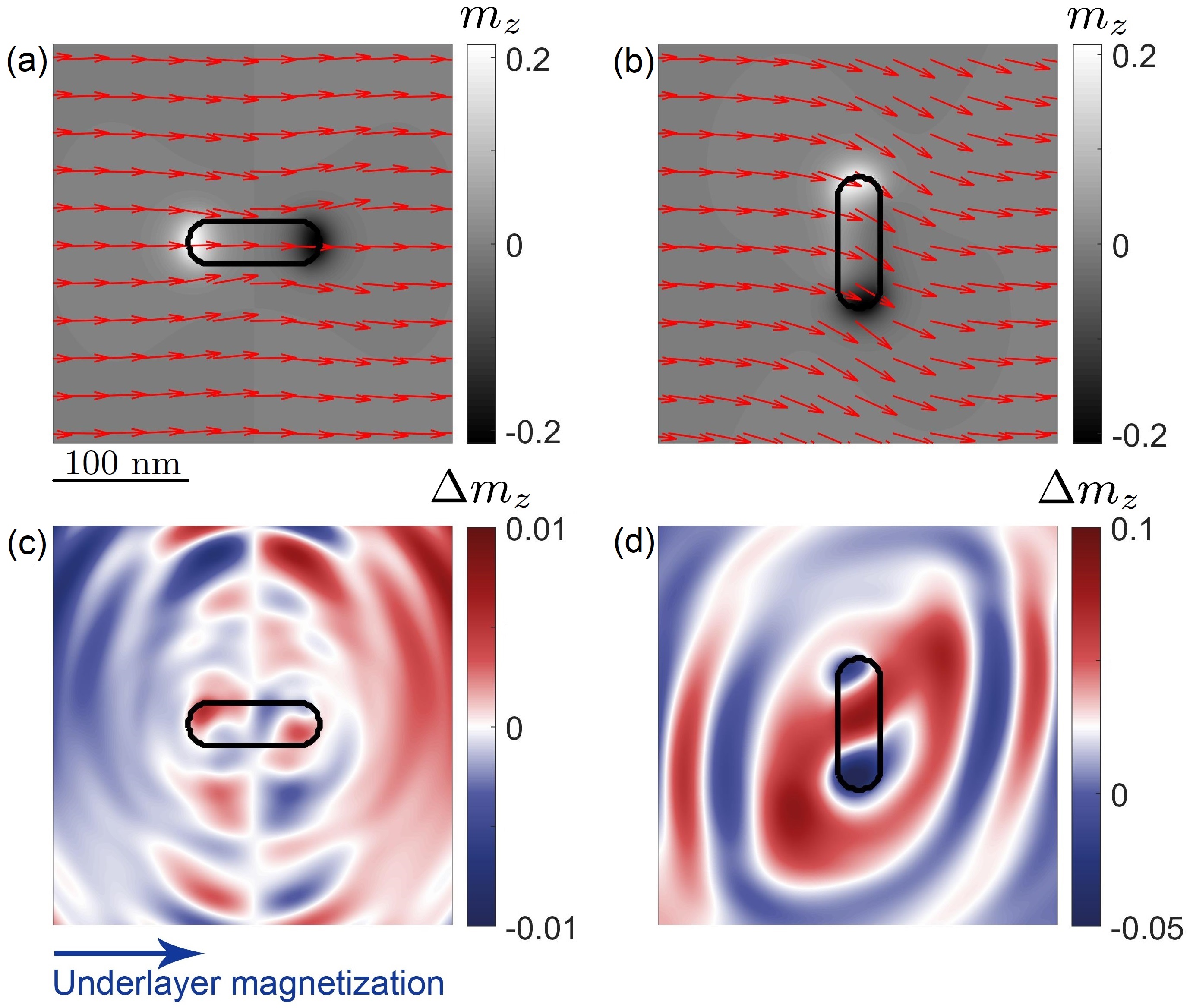}
\caption{ \label{fig2} Static magnetization resulting from coupling between the underlayer and a single nanoisland oriented (a) parallel and (b) perpendicular to the underlayer exchange bias. {The greyscale colormap indicates the {out-of}-plane magnetization component, $m_z$, and the arrows indicate the local in-plane direction of the magnetization.} A snapshot of the spin waves generated in the underlayer is shown for each case in (c) and (d), respectively, following application of a field pulse along the {($\bar{1}\bar{1}$)} direction. The snapshot is taken 50 ps after the field pulse. {The colorscale represents the spin-wave amplitude along the $z$ magnetization component, $\Delta m_z$.} }
\end{figure}
In this section, we consider a single nanoisland coupled to an extended soft magnetic underlayer. The underlayer is a Py film with thickness $10$~nm and the same nominal parameters as in Section A. The exchange bias~\cite{Nogues1999} is numerically modeled as an external field of $5$~mT along the $+x$ axis applied only at the bottom edge of the film, mimicking the interfacial nature of exchange bias. The 5-mT external field  corresponds to an interfacial exchange bias of 40~$\mu$J/m$^2$. This reduced exchange bias can be achieved in practice by dusting or inserting a thin spacer layer between an antiferromagnet, such as IrMn, and the Py underlayer~\cite{yanson2008,ali2008,zhao2015}. A Py nanoisland is placed at the geometric center of a circular underlayer with radius 480~nm. This geometry is chosen to minimize reflections at the boundaries. 
To avoid strong exchange interaction between the nanoisland and the underlayer, we assume a 90\% reduced exchange coupling, $A_\mathrm{interface}=1$~pJ/m. This can be realized in practice, e.g., via Cu or Ta  dusting of the underlayer-nanoisland interface~\cite{moskalenko2014,lee2009Ta}. While direct exchange is minimized, stray fields still provide a strong coupling mechanism between the nanoisland and the underlayer. 
{Simulation details are given in Appendix~\ref{app:spinwave}.}

The static configurations of the underlayer's magnetization in this system are shown in Fig.~\ref{fig2}(a) and (b) for, respectively, nanoislands with their easy axis parallel or perpendicular to the exchange bias direction. In the case illustrated in Fig.~\ref{fig2}(a), the parallel magnetization direction in both the nanoisland and the underlayer leads to {relatively} small changes in the underlayer in-plane magnetization, represented by red arrows. However, in the case illustrated in Fig.~\ref{fig2}(b), the coupling to the nanoisland induces a smooth non-uniform texture in the underlayer magnetization. In both cases, the magnetization tilts by ca. $12^\circ$ out-of-plane at the nanoisland edges, as quantified by the gray colormap.

In general, magnetization dynamics are perpendicular to the equilibrium magnetization. Therefore, we expect induced dynamics in the underlayer near the nanoisland to be anisotropic and to depend strongly on the relative orientation between the nanoisland and the underlayer equilibrium magnetizations. 
Magnetization dynamics are excited by applying a magnetic field pulse of amplitude $1$~mT and duration $50$~ps along the {($\bar{1}\bar{1}$)} direction. Snapshots of the out-of-plane component {deviation from the equilibrium  magnetization ($\Delta m_z$)} 50~ps after the pulse are shown in Fig.~\ref{fig2}(c) and (d): in both cases, the nanoislands act as spin wave sources. In Fig.~\ref{fig2}(c), excitation of the magnetization at the nanoisland edges leads to the emission of spin waves in the underlayer. The spin wave pattern is reminiscent of the electric field generated by a half-wave dipole antenna.
In contrast, in the case depicted in Fig.~\ref{fig2}(d), the distorted texture in the underlayer together with the orientation of the nanoisland magnetization give rise to anisotropic spin wave propagation, 
also originating in the vicinity of the nanoisland edges. {In addition, spin waves are more efficiently generated when the nanoisland's magnetization is perpendicular to that of the underlayer, as evidenced by the larger spin wave amplitude in Fig.~\ref{fig2}(d) compared to Fig.~\ref{fig2}(c)}. We expect that spin waves in the underlayer can propagate a large enough distance to mediate coupling between nanoislands in a lattice. Moreover, the very different nature of the propagation in the two illustrated cases clearly indicates that the coupling in {artificial} spin ice lattices will strongly depend on the local magnetization of individual nanoislands as well as the geometrical arrangement of the nanoislands. 
\begin{figure}[t]
\centering \includegraphics[width=3.2in]{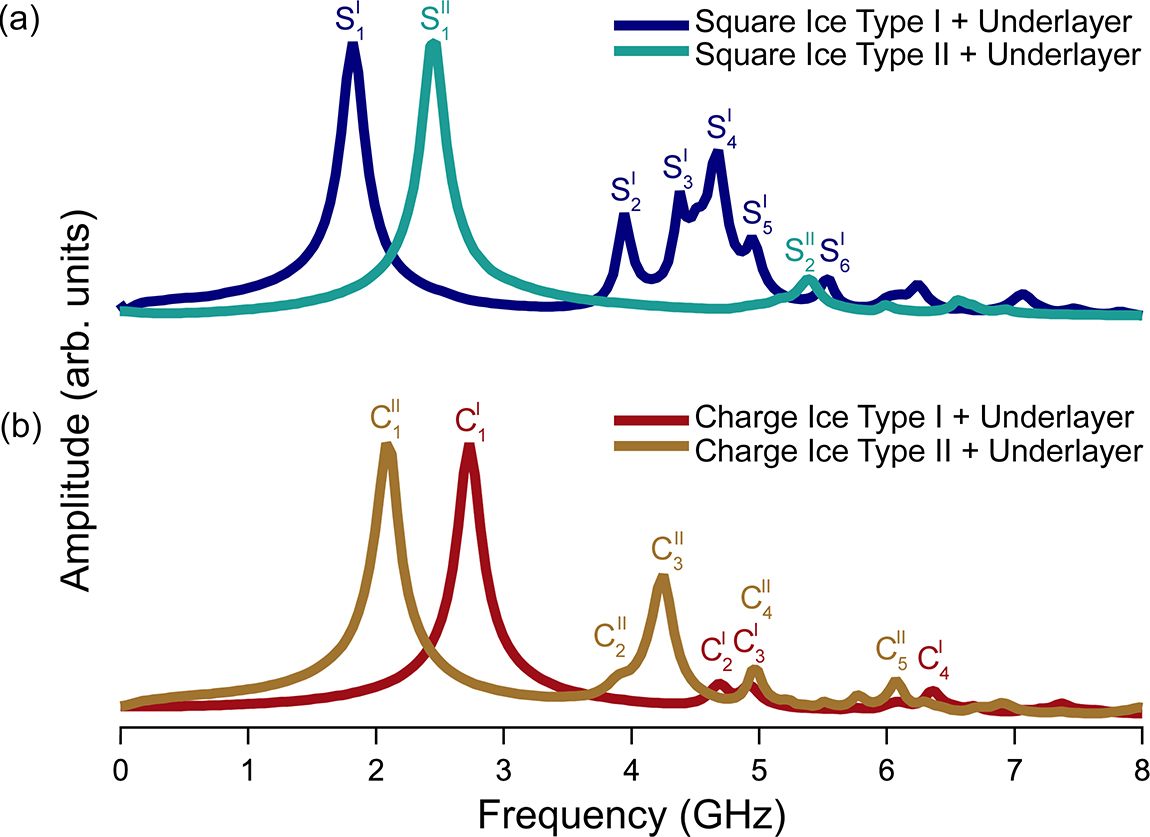}
\caption{ \label{fig5} Spectra {for} Type-I and Type-II configurations in (a) square ice and {in} (b) charge ice coupled to an exchange-biased underlayer.}
\end{figure}
\begin{figure*}[t]
\centering \includegraphics[width=7.0in]{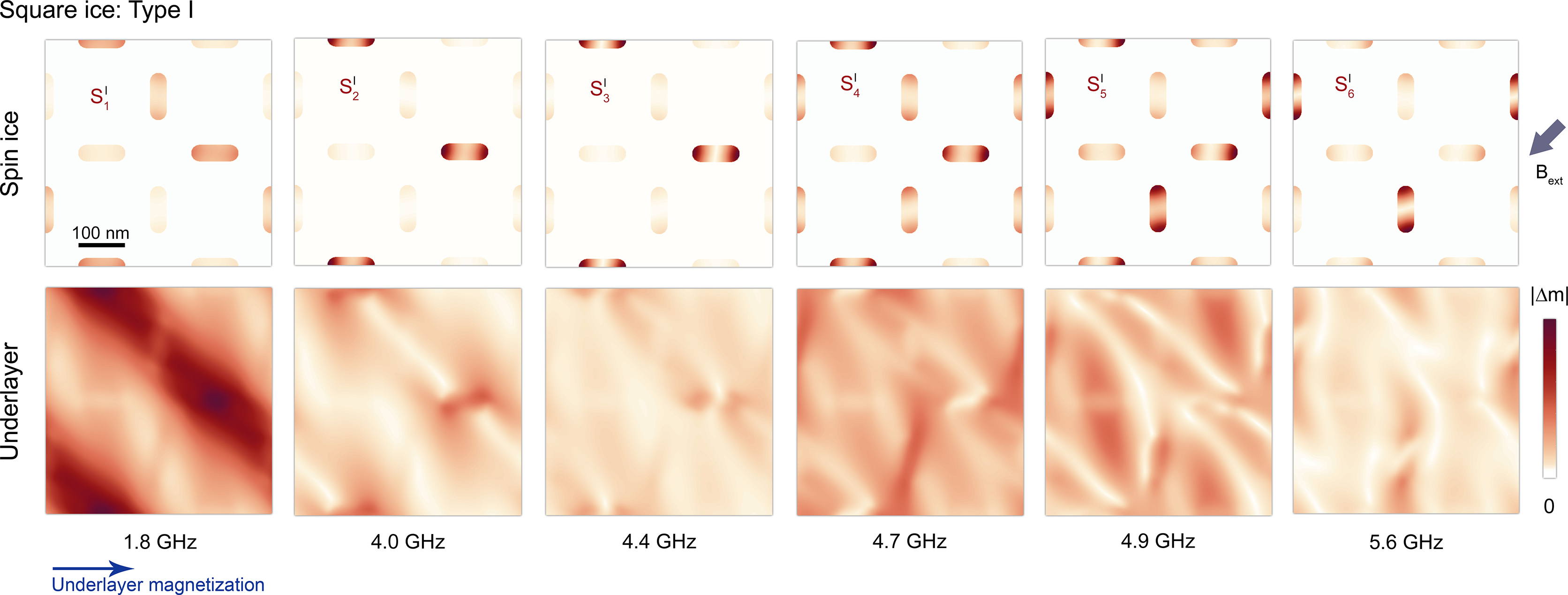}
\caption{ \label{fig:SI_Type_I} Spatially-resolved modes for the square ice Type-I configuration excited by a field pulse{, $B_\mathrm{ext}$,} along the ($\bar{1}\bar{1}$) direction, corresponding the spectrum in Fig. 5(a). The colormap represents the mode amplitude. The top row shows the mode distribution and amplitude in the spin ice, while the bottom row shows the mode distribution and amplitude in the underlayer. The direction of the exchange bias in the underlayer is indicated by the blue arrow.}
\end{figure*}

\subsection{{Dynamics of} artificial spin ices coupled to an exchange-biased underlayer\label{subsec:ASSI_MCSI}}

In this section we investigate the modes in {artificial} spin ices coupled to an underlayer. We follow the simulation procedure of section~\ref{subsec:square_and_charge} with the addition of an extended underlayer that is exchange-biased along the $+x$ axis and is exchange-coupled to the nanoislands with a reduced exchange constant of $A_\mathrm{interface} = 1$ pJ/m. Periodic boundary conditions are imposed in the $x-y$ plane for both the nanoislands and the underlayer.

\begin{figure*}[t]
\centering \includegraphics[width=6.2in]{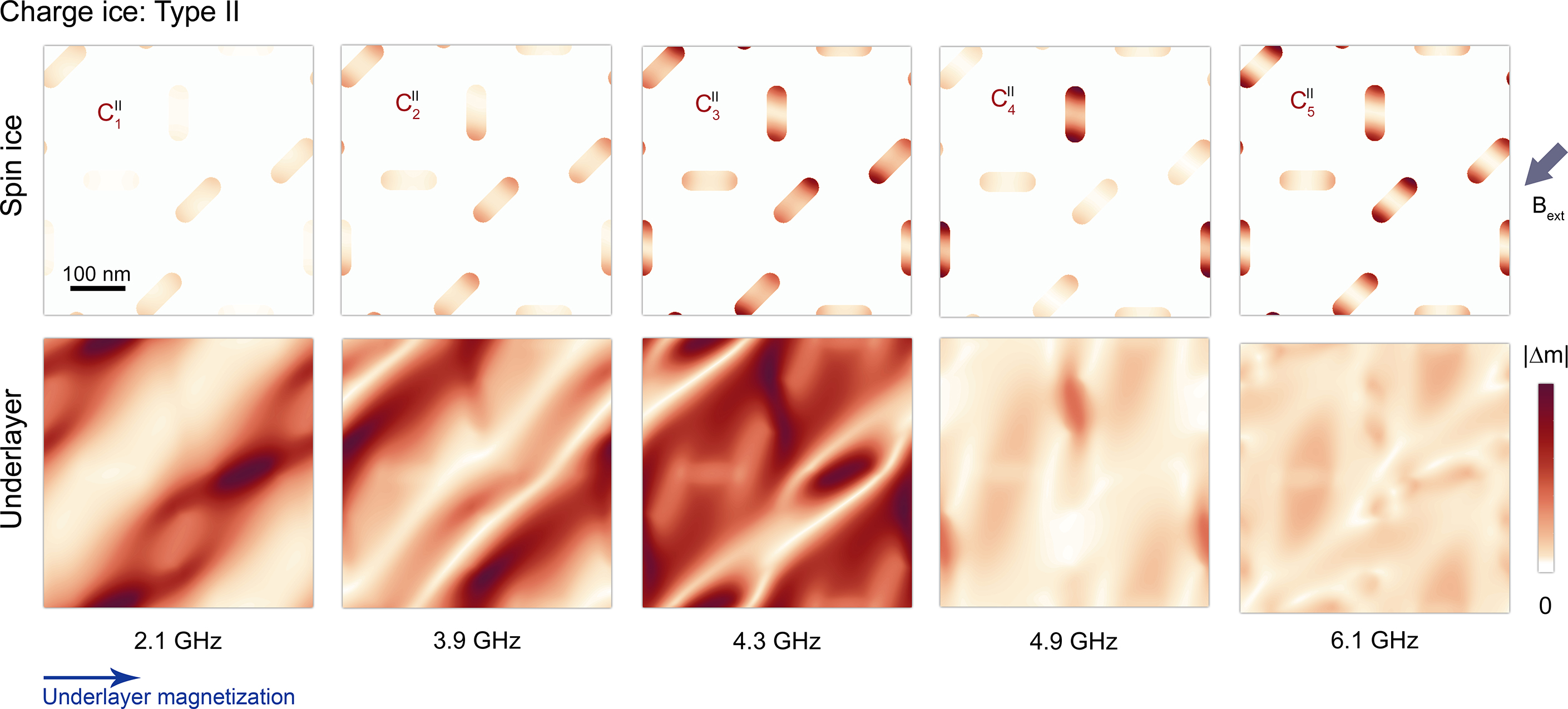}
\caption{ \label{fig:MC_Type_I} Spatially-resolved modes for the charge ice {Type-II} configuration excited by a field pulse{, $B_\mathrm{ext}$,} along the ($\bar{1}\bar{1}$) direction, corresponding the spectrum in Fig. 5(b). The colormap represents the mode amplitude. The top row shows the mode distribution and amplitude in the spin ice, while the bottom row shows the mode distribution and amplitude in the underlayer. The direction of the exchange bias in the underlayer is indicated by the blue arrow.}
\end{figure*}
The spectra of the coupled systems, are shown in Fig.~\ref{fig5}(a) and (b) for the square and charge ices, respectively. The inclusion of the underlayer leads to dramatic changes to the spectra when comparing to Fig.~\ref{fig:square_ice_I}(a). Clearly, the coupling to the underlayer lifts the degeneracy of the spectra for the Type-I and Type-II configurations. These spectra are characterized by relatively large shifts between the lowest frequency modes of Type-I and Type-II configurations for both the square and charge ices. These features are indicative of an increased interaction between the nanoislands mediated by the underlayer. {Interestingly, the high-frequency bulk mode (previously at ca. 12 GHz in Section IIA) in the nanoislands is quenched in all cases (not shown in Fig.~\ref{fig5}). This is due to the fact that spin waves in the underlayer are primarily excited through dipolar fields originating in the nanoislands, as shown in Fig.~\ref{fig2}. In other words, the bulk mode does not couple strongly to the underlayer resulting in effectively small-amplitude dynamics relative to the total magnetic volume of the system.}

Inspection of the spectra also shows that {one} configuration exhibits a richer spectral content {than the other} for both the square and charge ices. See, e.g., {the Type-I square ice} modes S$^I_2$-S$^I_6$ and {the Type-II charge ice modes C$^{II}_2$-C$^{II}_5$}. 
When the nanoislands in an artificial spin ice interact, degeneracies of non-interacting modes are lifted and give rise to a more complex spectrum (cf. Figs.~\ref{fig:square_ice_I} and \ref{fig5}). {For example, the Type-I square ice} configuration has four nanoislands in the unit cell {that each couples differently to the} exchange-biased underlayer. {Consequently,} the interactions between {the magnetization in the underlayer and in the nanoislands will give rise to a rich mode spectrum}. In contrast, {only two nanoislands form the unit cell of the square ice Type-II configuration,} so here the manifold of modes that arises from interactions will not be as rich as that of the Type-I configuration. {We also note that the Type-II square ice modes are qualitatively similar to the Type-I charge ice modes. As we show below, this similarity stems from the relative orientation between the nanoislands' magnetization to the underlayer.} To gain deeper insight, we investigate the spatially-resolved profile of the modes in the following sections.

\subsubsection{\label{sec:TypeI}{Square ice Type-I and charge ice Type-II} spatially-resolved modes}

The spatially-resolved modes of the Type-I square ice are shown in Fig.~\ref{fig:SI_Type_I} for both the nanoislands and the underlayer within the simulated {supercells}. The dominant mode S$_1^I$ is an edge mode of the nanoislands and the anisotropic propagation of the spin waves in the underlayer gives rise to spin wave channels perpendicular to the external field {pulse}. These channels connect horizontal and vertical nanoislands along every second diagonal of the {artificial} spin ice array (along the ($\bar{1}1)$ direction), thus providing an underlayer-mediated dynamical coupling between the nanoislands. Note that the power of this mode is mainly concentrated in the underlayer, indicating the robustness of these channels. The mode S$_2^I$ is similar in nature to S$_1^I$, however, its power is concentrated in the artificial spin ice while the spin wave channels in the underlayer are replaced by nodal lines linking horizontal nanoislands along every second diagonal. {The presence of nodal lines, instead of continuous spin wave channels indicates that{, while} the S$_1^I$ mode can support energy propagation in both the underlayer and artificial spin ice, the S$_2^I$ mode can only propagate energy through the artificial spin ice (through magnetostatic coupling). However, in both cases, the underlayer plays a fundamental role in increasing the coupling between the nanoislands.} Modes S$_3^I$ and S$_4^I$ are higher-order modes characterized by a large number of nodal lines within the underlayer as well as by the presence of nodal lines within the horizontal nanoislands {in} every second diagonal along the $(\bar{1}1)$ direction. The locations of nodal lines in the nanoislands suggest that S$_3^I$ is odd in the horizontal nanoislands. 
At higher frequencies, modes S$_5^I$ and S$_6^I$ are higher-order modes, which give rise to localized spin wave patterns ({instead of} 'patches') in the underlayer, while the power of these modes is concentrated in the spin ice.

The spatially-resolved modes for {Type-II} charge ice are shown in Fig.~\ref{fig:MC_Type_I}. Similar to the {Type-I} square ice, the dominant mode {C$_1^{II}$} exhibits diagonal spin wave channels in the underlayer. However, these channels are mediated by the diagonal nanoislands in the charge ice (which are parallel to the applied field pulse), providing a natural direction for coupled edge-mode excitations between the nanoislands and spin wave propagation in the underlayer. We note that the the power distribution along these channels is not uniform and that the strongest spin wave excitation in the underlayer occurs in regions between neighboring diagonal nanoislands, where constructive spin wave interference is achieved. {The modes {C$_2^{II}$ and C$_3^{II}$} are} in many respects similar to {C$_1^{II}$}, with additional channels between the diagonal and vertical nanoislands and with nodal lines along the direction of the spin {wave} channels found in {C$_1^{II}$}. The modes {C$_4^{II}$ and C$_5^{II}$} are higher-order modes localized both in the spin ice and the underlayer, where they exhibit a patched distribution.
\begin{figure*}[t]
\centering \includegraphics[width=6.5in]{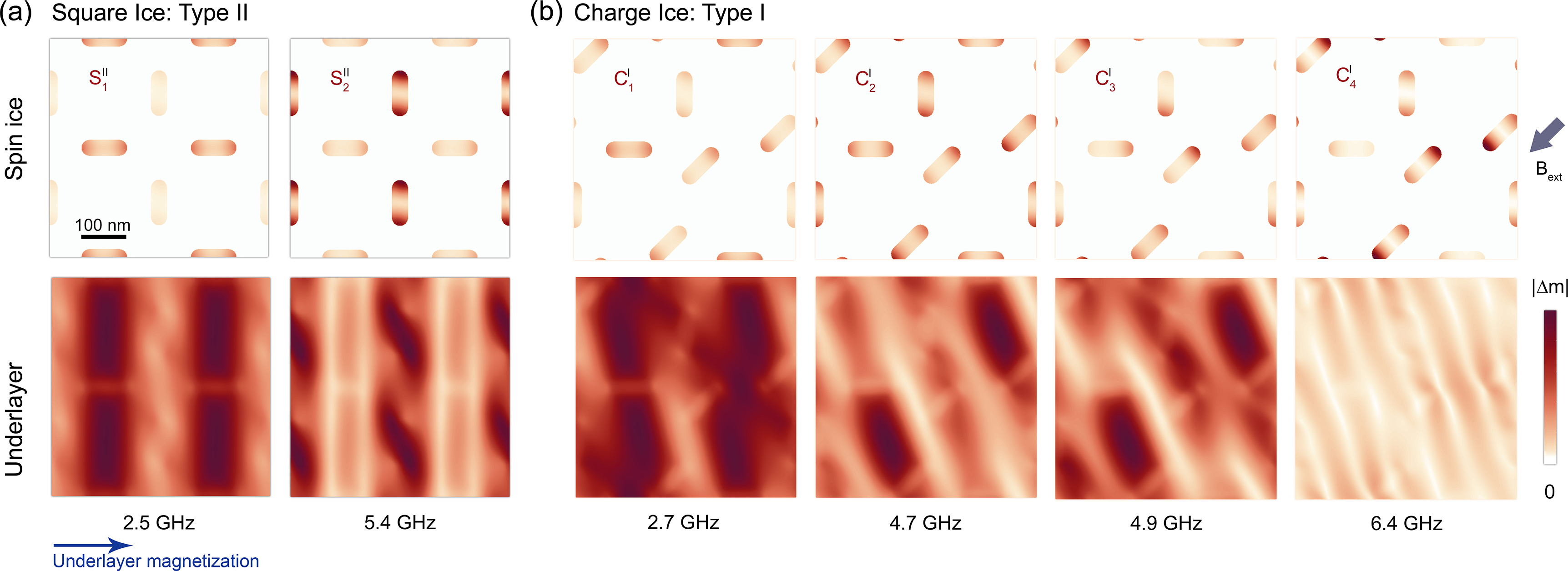}
\caption{ \label{fig:ASI_Type_I_II} Spatially-resolved modes for the {Type-II square ice} (a) and {Type-I charge ice} (b) magnetic configurations excited by a field pulse{, $B_\mathrm{ext}$,} along the ($\bar{1}\bar{1}$) direction. The modes correspond the spectra in Fig. 5(a-b). The colormap represents the mode amplitude. The top row shows the mode distribution and amplitude in the spin ice, while the bottom row shows the mode distribution and amplitude in the underlayer. The direction of the exchange bias in the underlayer is indicated by the blue arrow.}
\end{figure*}

\subsubsection{\label{sec:TypeII}{Square ice Type-II and charge ice Type-I} spatially-resolved modes}

For {the Type-II} square ice, the spatially-resolved modes for S$_1^{II}$ and S$_2^{II}$ are shown in Fig.~\ref{fig:ASI_Type_I_II}(a). There are pronounced differences between these modes and those of Type-I. Mode S$_1^{II}$ is associated to oscillations of the horizontal nanoislands which give rise to spin wave channels {along the (01) direction} connecting the edges of horizontal nanoislands. Oscillations of the vertical nanoislands have almost no amplitude. In contrast, mode S$_2^{II}$ is associated to oscillations of the vertical nanoislands and the mode power distribution in the underlayer is opposite of that in mode S$_1^{II}$: the vertical spin wave channels are replaced by nodes, while regions with low power in S$_1^{II}$ now display maxima, resulting in zig-zagging bands {originating} from the anisotropic excitation of spin waves from the nanoislands.

For the {Type-I} charge ice, shown in Fig.~\ref{fig:ASI_Type_I_II}(b), the dominant mode is also noticeably different from its {Type-II} counterpart. In {C$_1^{I}$}, most nanoislands oscillate establishing a more segmented spin wave pattern in the underlayer, which could nevertheless define spin wave channels along the {(01)} direction. While the power distribution of the {C$_2^{II}$} mode is primarily along the {(11)} direction {(See Fig.~\ref{fig:MC_Type_I})}, the power distribution {patterns of {C$_2^{I}$ and C$_3^{I}$} follow} a canted direction, connecting diagonal nanoislands to neighboring horizontal or vertical nanoislands. Higher-frequency modes correspond to preferential symmetries excited in the charge ice, leading to well-defined spin-wave patterns in the underlayer. Notably, for mode {C$_4^{I}$} the mode amplitude displays a large number of {nodes in the underlayer, resulting in a roughly periodic pattern of nodes along the centers of the diagonal nanoislands.}

\section{\label{sec:conclusion}Summary and conclusions}

We have demonstrated the emergence of complex dynamical behavior when {different} artificial spin ices are coupled to a soft magnetic underlayer, each of which exhibits relatively trivial dynamics on its own. {Using micromagnetic simulations, we established that the dynamic interaction between magnetic nanoislands and the underlayer enables spin-wave-mediated coupling between nanoislands located hundreds of nanometers apart.} While our modeling technically does not directly demonstrate dispersive propagating magnons, our results showing propagating dynamics in the underlayer, and coupled modes in the {artificial} spin ice-underlayer system, strongly suggest that these systems indeed support propagating magnons, and therefore comprise magnonic crystals. 
{The lowest-frequency modes are observed to support spin wave channels separated by nodes characterized by strongly suppressed or zero amplitude dynamics. For higher frequency modes, there are no spin wave channels, but instead, areas of localized dynamics connected by nodal lines or modes spatially delimited by the nodal lines. The} spin wave channels in the underlayer can effectively act as waveguides and can be reconfigured through the magnetic state of the superjacent spin ice lattice. In particular, the charge ice, which has been demonstrated to be {easily and reliably} reconfigurable using global magnetic fields, {could} allow for switching between distinctly different spin wave propagation directions as the magnetization state is toggled between {Type-I and Type-II} configurations. The experimental realization of the spin ice-underlayer system depends in practical terms on achieving a weak exchange bias on the underlayer~\cite{yanson2008,ali2008,zhao2015}, and a reduced exchange coupling between the nanoislands of the artificial spin ice and the underlayer~\cite{lee2009Ta,moskalenko2014}. {Optical techniques such as microfocused Brillouin Light Scattering ($\mu$BLS)~\cite{Sebastian2015} are most suitable to image the spatial modes. The lowest-energy modes have a spatial extent of approximately the lattice parameter of 290 nm, which is in principle measurable with $\mu$BLS.}

{The presented results} define a path to the realization of reconfigurable magnonic crystals with potential applications in  
spintronic and magnonic-based logic architectures~\cite{Chumak2017,Wagner2016}. These could enable simultaneous waveguiding and data processing through time-dependent spin-wave channel reconfiguration, enabling e.g. {alternative} schemes for computing~\cite{Jensen2018,Arava2019} using artificial spin ices.

{Input files for Mumax3 and data underlying the figures are available in the Materials Data Facility~\cite{Data,Blaiszik2016,Blaiszik2019}.}

\begin{acknowledgements}
E.I. and S.G. contributed equally to this work. S.G. was funded by the European Union's Horizon 2020 research and innovation programme under the Marie Sklodowska-Curie grant agreement no. 708674 {and by the Swiss National Science Foundation Spark Project Number 190736}. O.H. was funded by the US Department of Energy, Office of Science, Basic Energy Sciences Division of Materials Sciences and Engineering. Use of the Center for Nanoscale Materials, an Office of Science user facility, was supported by the US Department of Energy, Office of Science, Office of Basic Energy Sciences, under Contract No. DE-AC02-06CH11357. We gratefully acknowledge the computing resources provided on Bebop and Blues, high-performance computing clusters  operated by the Laboratory Computing Resource Center at Argonne National Laboratory. S.G. wishes to thank Michael Sternberg for support with the High-Performance Computing Cluster at the Center for Nanoscale Materials. 
\end{acknowledgements}

\appendix
\section{Micromagnetic simulations}
We used the Mumax3~\cite{Vansteenkiste2014} code set to utilize an adaptive-step, 4$^\mathrm{th}$ order Runge-Kutta algorithm. The temperature was set zero for all simulations. We have also used an in-house code~\cite{Heinonen2007} for cross-validation.

\section{Relaxation of the magnetic configurations}
\label{app:ground}

The simulations are initialized with a prescribed magnetization state (e.g., Type-I or Type-II with the magnetization in each nanoisland uniform along its length) and allowed to relax to its energy minimum by use of a combination of high damping $\alpha=1$ for 10 ns and a realistic damping of $\alpha=0.01$ for 20 ns. This approach {eliminates} ringing that might originate from the relaxation at low $\alpha$. The criterion for equilibrium was based on fluctuations of the magnetization vector on the order of single-precision.

\section{Spin ice and coupled systems eigenmodes}
\label{app:spinwave_ASI}

To accurately resolve the dynamics of a lattice, we discretize the simulated domains in cells of size 0.93~nm$\times$0.93~nm$\times$5~nm. The form factor of the cells has been shown to have negligible effect on the spectra~\cite{Iacocca2017c} while the small lateral size is required to accurately resolve the edge modes in the Py nanoislands~\cite{Gliga2013} in a finite differences approach. {Periodic boundary conditions are used in the $x-y$ plane for both nanoislands and underlayer to simulate an extended artificial spin ice lattice. Such boundary conditions guarantee that the micromagnetic supercells shown in Fig.~\ref{fig1} compose a faithful representation of the full artificial spin ice geometrical arrangement.} Once the magnetization has been relaxed following the procedure outlined in Appendix~\ref{app:ground}, the system is excited by applying an external field pulse $B_\mathrm{ext}=1$~mT for 50~ps along the ($\bar{1}\bar{1}$) direction. A sufficient frequency resolution of ca.~40~MHz is achieved by sampling the magnetization every 10~ps for 25~ns. By Fourier transforming the magnetization dynamics in each discretization cell, we obtain the spectra shown in Figs.~\ref{fig:square_ice_I} and \ref{fig5}. The spatial profile of each mode is obtained by selecting the frequency ranges of interest and transforming back from the frequency domain to the time domain.

\section{Single nanoisland spin wave excitation}
\label{app:spinwave}

In this case, the underlayer is assumed to be a disk with a radius of $480$~nm. {To minimize spin-wave reflections from the disk's edge, we incorporated absorbing boundary conditions in the form of a linear increase of the magnetic damping parameter. The absorbing region was a circular shell of width $200$~nm where the damping was increased at a rate of $0.5$~nm$^{-1}$. The simulated domain is discretized in cells of size 1.8~nm$\times$1.8~nm$\times$5~nm. This discretization implies that the absorbing boundary conditions are composed by approximately 110 concentric rings where the magnetic damping increases from its physical value up to a fictitious value, which is 100 times larger. Using these boundary conditions, the ground state is achieved by following the procedure outlined in Appendix~\ref{app:ground}}. Spin waves are excited by applying an external field pulse $B_\mathrm{ext}=1$~mT for 50~ps along the {($\bar{1}\bar{1}$)} direction. The spin wave evolution is simulated for a total of 100~ps with magnetization snapshots captured at each ps.

\end{document}